\documentstyle[preprint,aps]{revtex}
\begin{document}

\draft

\title{Absence of a Phase Transition in a Three--Dimensional Vortex
Glass Model with Screening}

\author{H. S. Bokil and A. P. Young}
\address{Department of Physics, University of California, Santa Cruz, CA 95064}

%\date{\today}

\maketitle

\begin{abstract}
We study the gauge glass model for the vortex glass transition in type--II
superconductors, including 
screening of the interaction between vortices. From the size dependence
of the domain wall energy
we find that, in two--dimensions,
the transition is at $T=0$ both with and without screening but the
exponents are different in the two cases. In three-dimensions,
we find that screening destroys the
finite temperature transition found earlier when screening was
neglected.

\end{abstract}


The question of whether the linear resistance, i.e. the resistance in the
limit of a vanishingly small current~\cite{linear-R}, is really zero in a
type--II superconductor in fields greater than $H_{c_1}$ has received
considerable attention recently, both theoretical and experimental,
because fluctuation effects~\cite{ffh}
are so much bigger for high
$T_c$ compounds than for low $T_c$ materials. In
fact much of the $H-T$ phase diagram of high $T_c$ materials is
occupied by a ``vortex liquid'' regime in which the resistance has
dropped because superconducting short range order has formed, so flux lines
exist locally, but the resistance is not yet zero because the flux
lines move under the action of a
Lorentz force due to the current, and hence give rise to a
voltage~\cite{ks}.
An important question, then, is whether, at lower--$T$,
the flux lines can be collectively pinned by
defects sufficiently strongly
that they have no linear response to the Lorentz
force, in which case the linear resistance is zero. Such as state was
first proposed for bulk superconductors by M.~P.~A.~Fisher~\cite{mpaf}
and called the vortex glass. Note that disorder, which is essential to
get a vanishing resistance, destroys the flux lattice~\cite{lo}
predicted by mean field theory.

It is important to distinguish whether the dominant type of disorder is
point--like, which could be caused by oxygen vacancies in high--$T_c$
materials, or whether it is due to extended defects such as twin
boundaries or artificially created columnar defects,~\cite{nv}.
The latter case is analogous to bosons
moving in a random potential in two dimensions~\cite{n}. For
this problem the vortex glass state is called a Bose glass~\cite{fwgf},
which undoubtedly occurs~\cite{bg_trans} in three dimensions, though
some controversies still remain about the details.
However,  the situation is less clear for point disorder, the case studied
here. Much of the theoretical work has
studied a simplified model, called the gauge glass.
%whose Hamiltonian is
%given in Eq.~(\ref{ham}) below.
In two dimensions, this has a transition
at $T = 0$~\cite{fty,other_dwrg},
whereas three dimensions seems close to the lower
critical dimension~\cite{rtyf,other_dwrg},
though there does appears to be a finite
$T_c$~\cite{rtyf} in this case. Several experiments~\cite{expts}
on samples of YBCO have found
reasonably convincing evidence for a phase transition in non-zero
field, presumably to some sort of vortex glass state.

Studies of the vortex glass transition
with point disorder have so far neglected 
screening of the (otherwise logarithmic) interaction between
vortices. As we shall see below, screening cannot be neglected very close
to $T_c$. The purpose of this paper is to see if the
finite--$T_c$ vortex glass transition survives the introduction of 
screening. We find, at least for the gauge glass model, that it
does not. Hence, assuming that the gauge glass model is in the same
universality class as experimental systems with point disorder, the 
observed ``transition'' in these systems would actually be rounded out
sufficiently close to $T_c$.
%Although this rounding will be difficult to see
%because it only occurs  very close to the transition, it may be
%possible to detect it in sensitive experiments on high--$T_c$ materials.
%**** REPHRASE. *****
%In fact experiments by **** using picovolt measuring techniques 
%find somewhat different results (scaling function) from those
%of **** which used larger voltages and hence were not so close to the
%transition (which is at zero voltage and current). It is possible that this 
%difference is due to the effects of screening, which would be more
%pronounced in the experiments of ****.

Screening will alter the critical behavior when the penetration depth,
$\lambda$, becomes equal to the correlation length, $\xi$, because the
important fluctuations on scale $\xi$ will no longer be screened. 
The behavior of $\lambda$ and $\xi$ are sketched in Figure 1. 
In the mean field regime, both $\lambda$ and $\xi$
diverge as $t^{-1/2}$, where $t= T - T_c$, so the ratio
$\lambda / \xi \equiv \kappa\ ( > 1)$ is roughly constant
and is a property of the material.
As the temperature is reduced there will be a crossover from mean field
behavior to the critical behavior of an uncharged superfluid, in which
screening can be neglected, at the Ginzburg temperature, $T_G$~\cite{ffh}.
%for bulk samples, occurs when~\cite{crit_phenom}
%$ \lambda(T_G)^2 / \xi(T_G) \sim \Lambda_T $,
%with $\Lambda_T = \phi_0^2 / 16 \pi^2 k_B T$, where
%$\phi_0 \equiv h c / (2 e)$ is the flux quantum.
In this critical regime, $\xi$ increases faster than
$\lambda$~\cite{ffh}, so
the two lengths eventually become equal at $T= T_{sc}$, say.
Screening effects will therefore be important for $T_c < T < T_{sc}$.
Since $\xi(T_{sc}) \sim \lambda(T_{sc}) \sim \Lambda_T$, we have
$\xi(T_{sc}) / \xi(T_G) \sim \kappa^2$, see Figure 1. 
Now $\kappa\simeq 100$ for
high $T_c$ compounds so the region where screening is important is
small and has probably not been accessed in experiments so far.
Nonetheless we feel that it is
useful to study the behavior in this region, because (i) this regime
may be accessible in experiments on certain materials and ranges of
field, and (ii) theoretically, it is interesting to know whether or not the
linear resistance is really zero.
%Taken literally this result suggests that screening will be important
%only in a region extremely close to $T_c$. In
%practice, numerical factors and lattice anisotropy effects
%(particularly strong for the Bismuth compounds, which are almost
%two-dimensional) will increase the size of the region.

Next let us discuss the gauge glass which we model used to study effects of
screening. Including a fluctuating vector potential,
the Hamiltonian is
\begin{equation}
{\cal H} = - \sum_{\langle I,J\rangle}
\cos ( \phi_I - \phi_J - A_{IJ} - e {\cal A}_{IJ})
 + {1\over 2} \sum_\Box \left(\vec\nabla \times
\vec{\cal A} \right)^2  \quad .
\label{ham}
\end{equation}
The phase of the condensate,
$\phi_I$, is defined on each site, $I$, of a regular lattice,
square for two dimensions, and simple cubic for $d = 3$,
with $N = L^d$ sites.  For now we assume periodic boundary conditions.
The sum is over all nearest neighbor pairs on the
lattice. The effects of the external magnetic field and disorder are
represented by the quenched vector potentials, $A_{IJ}$, which we take
to be independent random variables with a uniform distribution
between 0 and $2\pi$. The fluctuating vector potential on each link,
${\cal A}_{IJ}$, is integrated over from $-\infty$ to $\infty$~\cite{gauge}.
In the last term,
which is the usual magnetic energy, the sum is over all elementary squares on
the lattice,
and the curl, $\vec\nabla \times \vec{\cal A}$,
is the directed sum of the
vector potentials round the square, taking into account that ${\cal A}_{IJ} 
= - {\cal A}_{JI}$. The ``charge'', $e$, is a measure of the coupling between
the condensate and
the fluctuating vector potential.
The previously
studied model~\cite{fty,rtyf}, which neglects screening, corresponds to
$e=0$.

It is technically easier to study this model in the vortex
representation rather than the above phase representation. To do this we
replace the cosine by the periodic Gaussian (Villain)
function~\cite{villain}, and perform
standard manipulations~\cite{villain,vortex_model}. For
our purposes it is also essential to incorporate the periodic boundary
conditions, which leads to an additional term~\cite{fty,mpaf_priv_comm}.
For two dimensions, we find that the vortex Hamiltonian is
\begin{equation}
\label{vortex_2d}
{\cal H}_V = -{1\over 2} \sum_{i,j}
(n_i - b_i)G(i - j) (n_j - b_j) +
\sum_{\alpha= x,y} V(\Phi^\alpha - \epsilon^{\alpha\beta}P^\beta) \quad ,
%V(\Phi^x - P^y) + V(\Phi^y + P^x) \quad ,
\end{equation}
where the $\{n_i\}$ run over all integer values, subject to the
``charge neutrality'' constraint $\sum_i n_i = 0$, and are interpreted as
the vortices, $G(i - j)$ is the screened vortex interaction, 
\begin{equation}
\label{vortex_int_2d}
{G(i - j) \over (2 \pi)^2} = 
{1 \over N } \sum_{{\bf k} \ne 0}
{1 - \exp[i {\bf k} \cdot ({\bf r_i} - {\bf r_j})] \over
4 - 2 \cos k_x - 2 \cos k_y + e^2} \quad,
\end{equation}
and $V(u)$ is the Villain potential given by
\begin{equation}
\exp\left[{-V(u) \over T}\right] =
\sum_{m=-\infty}^\infty \exp\left[{-{1\over 2T} (u - 2 \pi m)^2}\right] \quad .
\end{equation}
$V(u)$ is periodic with period $2 \pi$, and, at $T=0$, is equal to
$u^2/2$ for $-\pi < u < \pi$.
The vortices sit on the sites, $i$, of the dual lattice
which are in the centers of the squares of the original lattice.
The $b_i$ are given by (1/$2 \pi$) times
the directed sum of the quenched vector potentials on
the links of the original lattice which surround the site $i$ of the
dual lattice. 
As usual
$\epsilon^{xy} = - \epsilon^{yx} = 1, \epsilon^{xx} = \epsilon^{yy} = 0.$
The function $\Phi^x$ is defined by
\begin{equation}
\label{phidef}
\Phi^x = \sum_x A_x \left(x + {1\over 2}, {1\over 2}\right) \quad ,
\end{equation}
and similarly for $\Phi^y$. Here we indicate explicitly the $x$
and $y$ coordinates of the lattice points: we use integers for the dual
lattice, $x = 1,2, \ldots, L$, so the coordinates of the original
lattice are half--integers.
The link variable $A_x (x + 1/ 2, 1/ 2)$ connects sites $(x+1/2,1/2)$
and $(x+3/2,1/2)$.
The function $\bf P$ is $2 \pi / L$ times the ``dipole moment'' of the vortex
and flux distribution, i.e.
\begin{equation}
{\bf P} = {2 \pi \over L} \sum_{\bf r} {\bf r}\ [n({\bf r}) - b({\bf r})]
\quad .
\end{equation}
The Hamiltonian is
independent of the choice of origin for the coordinate system: choosing
a new origin alters $\bf \Phi$ and $\bf P$ but it is straightforward
to show that the arguments of the Villain functions only change by a
multiple of $2 \pi$ and so the functions themselves are unchanged. 
Notice that for $e = 0$ the interaction between vortices is logarithmic
but that for $e \ne 0$ this logarithmic variation is screened beyond a
distance of order $1/e$.
%Note also that a
%change in the boundary conditions, say from
%periodic to anti-periodic in the $x$ direction, changes $\Phi^x$ by $\pi$,
%since this can be incorporated by adding $\pi / L$ to all the $A_x$. From
%Eq.~(\ref{vortex_2d}) one sees that the change in the energy of the
%Villain term due to
%this twist can be compensated for by
%moving a vortex through a distance $L/2$ in the $y$ direction (or
%moving many vortices with the same total change in the dipole-moment),
%though at the price of changing the vortex part of the energy, the
%first term in Eq.~(\ref{vortex_2d}).

In $d = 3$ the vortices are integer variables which lie on the links
of the dual lattice~\cite{vortex_model}.
We find that the vortex Hamiltonian, including
periodic boundary conditions, is
\begin{equation}
\label{vortex_3d}
{\cal H}_V = - {1\over 2}\sum_{i,j}
({\bf n}_i - {\bf b}_i) \cdot ({\bf n}_j - {\bf b}_j)\ G(i - j) +
\sum_{\alpha = x,y,z}
V\left(\Phi^\alpha -{1\over 2} \epsilon_{\alpha\beta\gamma}
C^{\beta\gamma} \right) \quad ,
\end{equation}
where the ${\bf n}_i$ are subject both to the global constraint
$\sum_i {\bf n}_i = 0$, analogous to $d=2$, and to the local constraint,
$[{\bf \nabla} \cdot {\bf n}]_i = 0$ for each site~\cite{vortex_model}.
The vortex interaction, $G(i-j)$, is obtained by replacing the RHS of
Eq. (\ref{vortex_int_2d}) by its
three-dimensional analogue, $C$ is given by
\begin{equation}
\label{cdef}
C^{xy} =  {2 \pi \over L^2} \sum_{\bf r} {1\over 2}
\left[x q^y({\bf r}) - y q^x({\bf r})\right]
+ {2 \pi \over L} \sum_{\bf r} {1\over 2}  \left[x q^y({\bf r}) \delta_{y,L}
- y q^x({\bf r}) \delta_{x,L}\right] \quad ,
\end{equation}
where ${\bf q}({\bf r}) = {\bf n}({\bf r}) - {\bf b}({\bf r})$, with
similar expressions for the other components, and
${\bf \Phi}$ is defined as in Eq.~(\ref{phidef}).
The physical interpretation of $C$ is quite simple. Consider the change
in $C$ when 1 is added to all the ${\bf n}_i$ on a closed loop in the
$xy$ plane. 
Eq.~(\ref{cdef}) tells us that $C^{xy}$ is then increased by
$2 \pi / L^2$ times the area of the
loop, while $C^{yz}$ and $C^{zx}$ are unchanged. It is easy to see
that this is true if the loop does not go along links~\cite{links} 
%$x=L+1/2$ or $y=L+1/2$,
at which there is a discontinuity in our labeling of the coordinates of
the sites,
because only the first term in Eq.~(\ref{cdef}) then contributes,
but it is not complicated to show that the result is true for all
loops. 
%If, however, the loop includes a link variable, such as $n^y(x,L,z)$, 
%which connects sites
%$(x, L, z)$
%and $(x, 1, z)$)
%where $y$ or $x$ decrease by $L$ rather than increase by 1,
%the second term in Eq~(\ref{cdef}) now contributes and the first term
%is no longer the area. Nonetheless, it is
%straightforward to show that the {\em total} expression for
%$C^{xy}$ is still equal to the area of the loop.
A change in the boundary conditions, from, say, periodic to
anti-periodic in in the $z$-direction, can be incorporated by adding
$\pi / L$ to all the $A_z$, and so has the effect of changing $\Phi^z$ by
$\pi$.
This can be compensated for, in the arguments of the Villain
functions, by forming a vortex loop in the $xy$ plane whose area is
$L^2/2$, or a combination of loops with the same total area.
A derivation of Eqs.~(\ref{vortex_2d}) and (\ref{vortex_3d})
will be given elsewhere~\cite{long}.

The gauge glass model has random fluxes with no preferred direction,
as distinct from the net applied field which occurs in experimental systems.
This difference is irrelevant in an expansion in $6-\epsilon$
dimensions, and so may also be irrelevant in $d=3$, though a firm
demonstration of this is lacking.
%It would obviously be useful to
%perform similar calculations on a more realistic model
%with a net field, though the
%anisotropy would make a finite size scaling analysis more difficult.

To investigate whether a transition occurs we calculate the defect wall
energy~\cite{defect_wall}. For a given realization of the disorder, we
compute the difference in ground state energy, $\Delta E$,
between the system with periodic boundary
conditions, and with boundary conditions which are anti-periodic in one
direction and periodic in the others. To find the ground state, we
start from different random initial configurations and quench down to
the nearest local energy minimum. We repeat this many times,
keeping track of the lowest energy found so far.
We monitor the 
r.m.s. energy difference, $\Delta E_{rms}$,
discussed below, after a logarithmically
increasing number of quenches, $1,3,10, \ldots$ and stop when the last
two estimates are virtually identical.
For a number of cases we ran for much longer than
this and verified that the results were unchanged. 
We also also performed some calculations in the phase representation,
using the Villain form for the interaction, and checked that identical
results were obtained.

For frustrated systems, such as the
model we study here, the average of $\Delta E$ over all samples will be zero,
so we look at the root mean square average,
$ \Delta E_{rms} = [ \Delta E^2]^{1/2}_{av}\ $, 
where $[\ldots]_{av}$ denotes an average over samples. We investigate the
size dependence of $\Delta E_{rms}$ and define an exponent $\theta$ by
\begin{equation}
\Delta E_{rms} \sim L^\theta \quad .
\end{equation}
If $\theta < 0$ then large domains cost little energy so, at any
finite $T$, the system will break up into domains and long range order
will be destroyed. By equating the domain wall energy on the scale of
the correlation length with $k_B T$ one
finds that the correlation length varies with $T$ as $T^{-\nu}$ with
$\nu = 1/|\theta|$. If, on the other hand, $\theta > 0$ then then the
rigidity of the ordered state should persist to finite temperature.

Our results for $d=2$ are presented in Figure 2 for $2 < L < 8$.
It was not possible to get convergent results for $\Delta E_{rms}$ for
larger sizes. For $e=0$ the slope is
$\theta \simeq -0.5$,
which is in good agreement with earlier work~\cite{fty}, and
gives a correlation length diverging as roughly $1/T^2$. Experimental evidence
for this has recently
been obtained~\cite{expt_2d} on very thin films of
YBCO. For $e \ne 0$ the slope is clearly more negative, indicating that
screening is a relevant perturbation.

The curvature of the data indicates that
the intermediate values of $e$ are in a crossover regime.
It is therefore be useful to
study directly the large $e$ limit. To do this, notice
from Eq.~(\ref{vortex_int_2d}) that the vortex energy becomes very small in
this limit, so the Villain term in Eq.~(\ref{vortex_2d}) dominates and
acts as a constraint fixing the dipole moments of the vortex
distribution to the values which minimize the Villain term. Also, in this
limit, $G(0)=0$, and $G(i-j)= (2 \pi / e)^2 $ plus exponentially small
corrections which we neglect. Because of charge neutrality, one can
subtract the constant $ (2 \pi / e)^2 $ from all the interactions, 
with the result that
only $G(0)\ (= -  (2 \pi / e)^2 )$ is non-zero.
Hence, for $e \to \infty$,
$\Delta E_{rms} e^2$ tends to a finite value which can be found by
by neglecting the Villain terms in the
energy, setting $G(0) = -(2 \pi)^2$ and $\ G(i-j) = 0$ for $i \ne j$, 
and performing a constrained minimization fixing not only the
total ``charge'' but also the total dipole moment of the vortex
distribution. On changing the boundary conditions, the 
$x$-component of the dipole moment is changed by $L/2$~\cite{even_odd}.
Results for  $e^2 \Delta E_{rms}$ obtained in this way
are shown in the inset to Figure 2.
The data follows a straight line on a log-log
plot with a slope of $\theta \simeq -1.36$,
which is also roughly the slope of the
data for the larger values of $e$ and $L$ in the main part of the Figure. 

This result is surprising because there are simple
arguments~\cite{hf_priv_comm} which predict that $\theta = -2$ in the
large $e$ limit. One way of seeing this is that, since the vortices don't
interact, the probability that a vortex is excited out of its ground
state is obtained just by comparing its on-site excitation energy
with $k_B T$. Since there is a finite density of states, the
density of excited vortices is of order $k_B T$, and hence a typical
separation (which we interpret to be the correlation length) of 
order $T^{-1/2}$, and so $\theta = -1/\nu = 2$. Assuming that this
is correct, the different result found here must reflect the
importance that the constraints (total charge = 0, fixed dipole moment)
have for the rather small system sizes studied here. Note, though, that
the presumably correct result, $\theta = -2$, predicts an 
even {\em faster} drop of
$\Delta E_{rms}$ than found in the numerics, and that the numerics is
{\em correct} in predicting that the screened system is in a
different universality class with a {\em weaker} divergence at $T=0$.

Our results for $\Delta E_{rms}$ in $d=3$ are
presented in Figure 3. The range of sizes, $L = 2, 3$ and 4, is very
small, but analogous calculations on similarly small sizes provided one
of the first clear indications for the, apparently correct, result that
$T_c$ is finite in 3-$d$ Ising spin glasses~\cite{defect_wall}. 
For $e = 0$, $\Delta E_{rms}$ is roughly independent of
size, in agreement with earlier work~\cite{rtyf}, which
indicates that the system is close to the lower critical dimension. 
For $e \ne 0$, however, $\Delta E_{rms}$, decreases with increasing size,
this effect being more pronounced for larger values of $e$. One can
also study the large $e$ limit, as for $d=2$, but we were only able to
get convergence for $L=2$ and 3,
and, with only two points, one cannot check for straight line behavior of
the log-log plot.
Although the sizes are very small,
the results of Figure 3 clearly imply that there is
no finite temperature transition in $d=3$.
The finite temperature transition, for which some evidence was
presented in Ref.~\cite{rtyf}, is therefore rounded out when one
includes screening.

To conclude, the presumed finite $T_c$ vortex glass transition in the
three--dimensional gauge glass model is rounded out by screening
effects. The same will be true in real
experimental systems with point disorder, 
assuming that the model is in the same universality class, which is
plausible but not yet firmly established.
Strictly speaking, the linear resistance would then
not vanish, though it would become extremely small because the rounding
only occurs very close to $T_c$.
It would obviously be interesting to investigate this model by finite
temperature Monte Carlo simulations, where larger sizes can be studied,
and also to simulate a more realistic model with a net field in a
particular direction.

We have benefited enormously from stimulating discussions with
M.~P.~A.~Fisher.
We would also like to thank D.~A.~Huse for valuable comments.
This work is supported by NSF grants DMR 91-11576 and 94-11964.

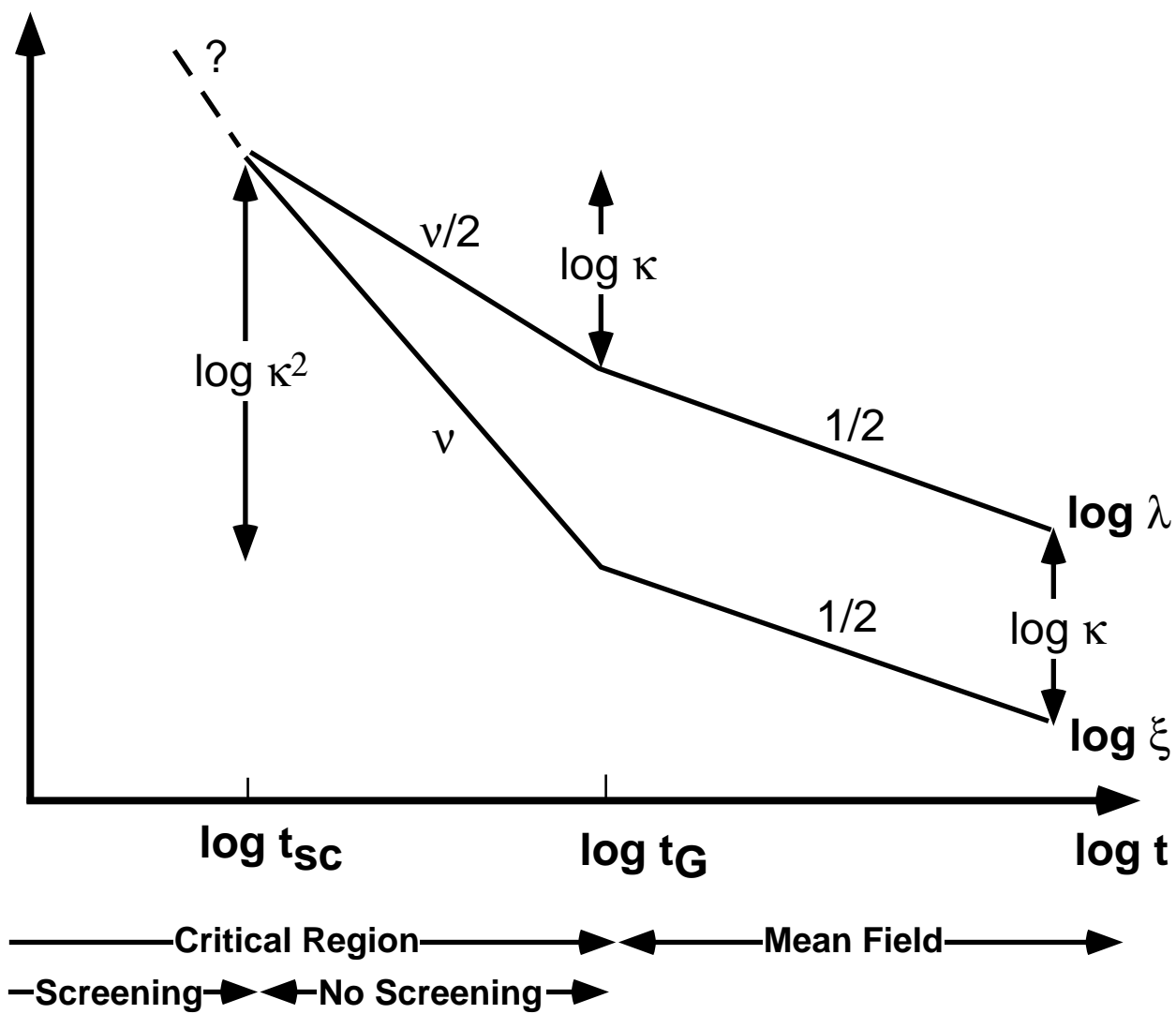
\begin{figure}
\caption{
A sketch of the log of the correlation length, $\xi$,
and penetration depth, $\lambda$, against $\log t$, where $t = T - T_c$,
for a type--II superconductor. The slopes of the curves are indicated:
$\nu$ is the correlation length exponent in the part of the
critical region where screening can be neglected. Two
crossovers occur. The first, at the reduced Ginzburg temperature,
$t_G = T_G - T_c$ is when
critical fluctuations first start to be important. The second, at 
$t_{sc} = T_{sc} - T_c$,
is when $\lambda$ and $\xi$ become comparable. Screening
effects are important for $t \le t_{sc}$.
For $T > T_G$, $\lambda / \xi  = \kappa\ ( > 1)$, is
constant, while for $T_{sc} < T < T_G$, $\lambda^2/ \xi$ is
constant~\protect\cite{ffh}. 
Between $T_G$ and $T_{sc},\ \xi$ increases by a factor of
$\kappa^2$ and $\lambda$ by a factor of $\kappa$.
}
\end{figure}

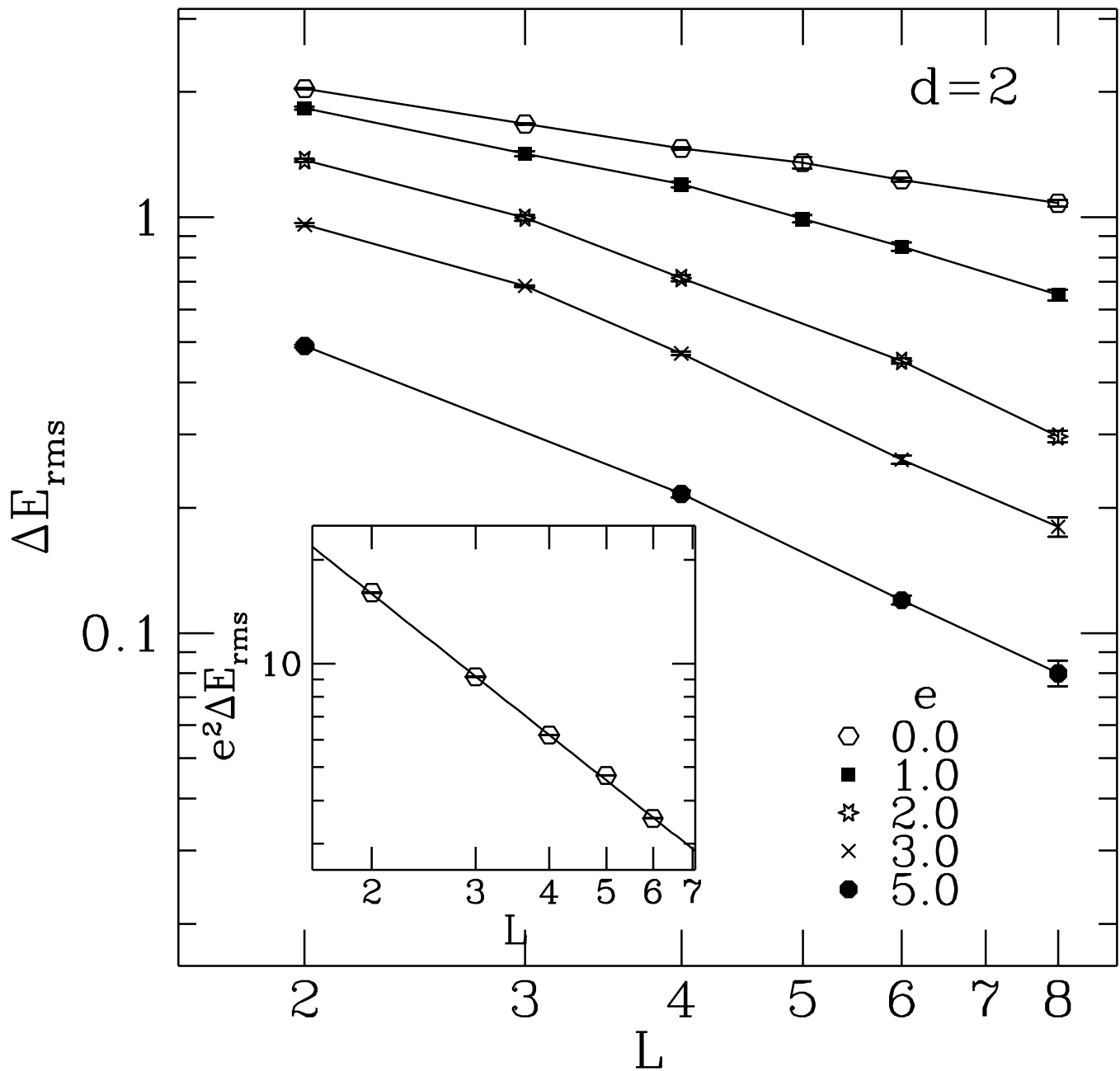
\begin{figure}
\caption{
A log--log plot of the r.m.s. domain wall energy,
$\Delta E$, against $L$ for $d = 2$ for
different values of the coupling $e$. The number of samples
varied between 4000 for the smallest  sizes to 200 for the largest sizes
and values of $e$. For $e=0$ the slope,
$\theta$, is approximately
$-0.5$ in agreement with earlier work~\protect\cite{fty}.
For $e \ne 0$ the curves bend down more steeply indicating that
screening is a relevant perturbation. 
The insert shows a
log--log plot of $e^2 \Delta E$ against $L$ for $d = 2$ 
in the limit $e \to \infty$. The number of samples
varied between 2000 for $L=2$ and 500 for $L=6$. The line is a least squares
fit and has a slope of $-1.36$.
}
\end{figure}

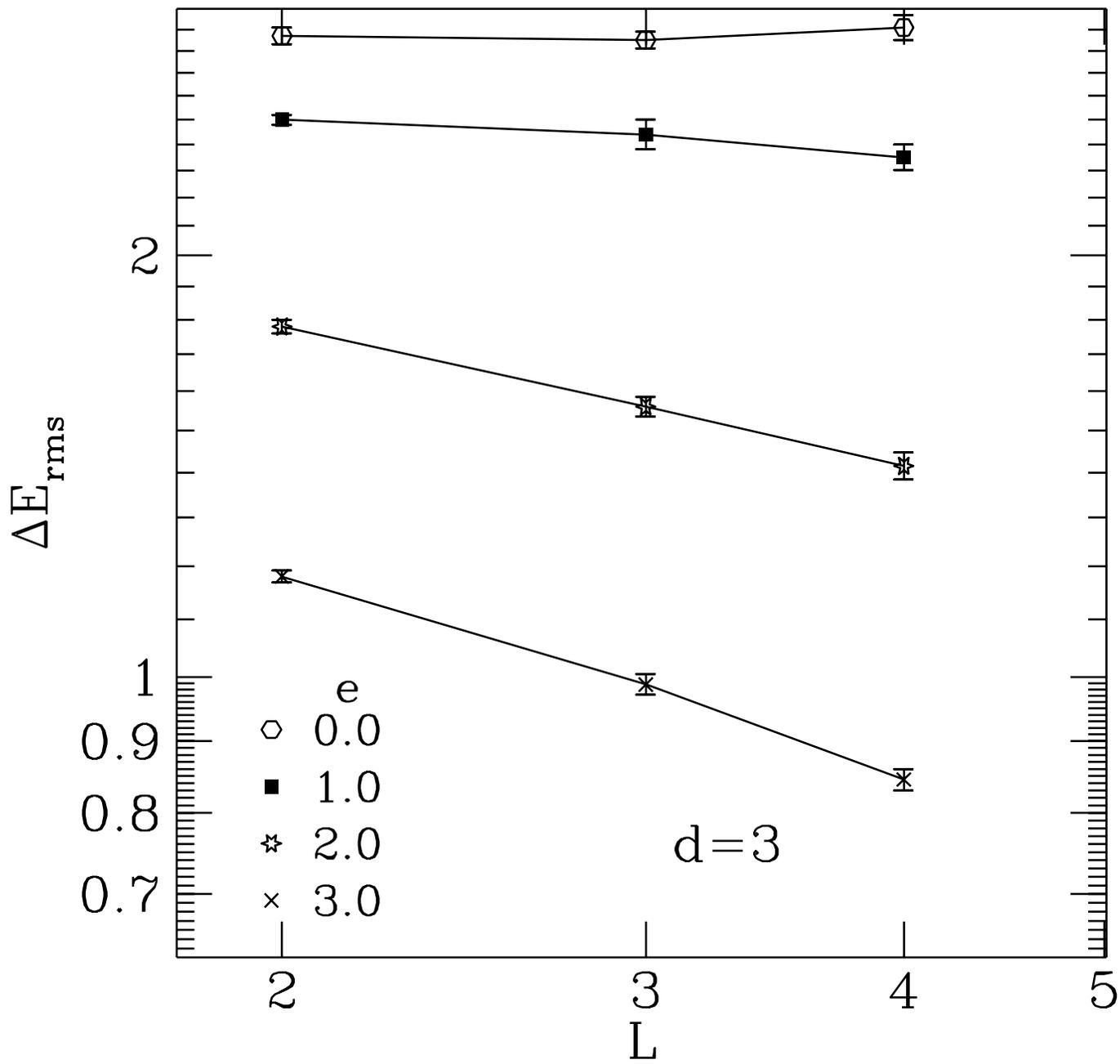
\begin{figure}
\caption{A log-log plot of $\Delta E$ against $L$ for $d = 3$ for
different values of the coupling $e$. The number of samples
was 4000 for $L=2$ and at least 1000 for $L=4$. For $e=0$ the slope
is close to zero, in agreement with earlier work~\protect\cite{rtyf}.
For $e \ne 0$ the curves bend down indicating that
screening is a relevant perturbation which destroys the (possible)
finite temperature transition at $e = 0$. 
}
\end{figure}

\begin{references}
\bibitem{even_odd} 
This also works for $L$ odd if one takes, for $L/2$, the nearest
integer. At large but {\em finite} $e$, where the Villain function part
of ${\cal H}_V$ is included, there is a strong even--odd effect,
because, for $L$ odd, moving the
vortices cannot precisely compensate for the change of $\Phi^x$ by $\pi$.
This is why only data for even $L$ are
presented in the main part of Figure 2.
\end{references}
\end{document}